**Magneto-optical magnetoelectric voltage sensor**

*Michael P. Path, Jeffrey McCord**

M. P. Path, J. McCord
Nanoscale Magnetic Materials - Magnetic Domains, Department of Materials Science, Kiel University, 24143 Kiel, Germany

J. McCord
Kiel Nano Surface and Interface Science (KiNSIS), Kiel University, 24143 Kiel, Germany

E-mail: jmc@tf.uni-kiel.de

Funding: German Research Foundation (DFG) - MC 9/20-2

Applications in high-voltage and electromagnetically harsh environments require reliable galvanically isolated voltage sensing, which can be achieved using optical readout. While established optical voltage sensors rely on electro-optic effects or piezoelectric strain with direct optical detection, strain-mediated magnetoelectric coupling combined with magneto-optical readout offers an alternative voltage sensing principle that remains largely unexplored. Here, such a sensor based on a bismuth-substituted yttrium iron garnet magneto-optical indicator film mechanically coupled to a piezoelectric actuator is presented. Voltage induced stress results in changes of the out-of-plane magnetization via magnetoelastic coupling which is detected through magneto-optical Faraday rotation. A critical state of the domain structure is set via an applied bias field, in which voltage-induced nucleation and domain-wall motion dominates the response. In this high sensitivity regime, both AC and DC voltage readout modes are demonstrated, based on either voltage-driven magnetization reversal or voltage-induced modifications of the magnetization loop shape. Equivalent voltage noise densities in the millivolt per root hertz range are achieved. The results establish strain-mediated magneto-optical voltage sensing as a distinct approach to optically isolated voltage measurement.

## 1. Introduction

Conventional voltage measurements rely on analog-to-digital converters preceded by voltage scaling and conditioning stages, such as inductive voltage transformers and resistive or capacitive dividers, which translate the voltage under test into a measurable signal level.[1] However, for the relevant high-voltage grid applications and their electromagnetically harsh environmental conditions, these approaches face limitations related to galvanic isolation, parasitic coupling, and electromagnetic interference.[2] Optical voltage sensors provide an alternative voltage transformation with intrinsic galvanic isolation and electromagnetic immunity.[2,3] Commercialized concepts are based on electro-optic (Kerr or Pockels) effect integrated into fiber-optics.[4,5] Optically inferring voltage using piezoelectric strain transducers, e.g. via fiber-Bragg-grating, has been proposed.[6–8]

Strain-mediated magnetoelectric (ME) coupling offers an additional route for voltage sensing, which is lacking thorough investigation. In such composites, an applied voltage produces piezoelectric strain, which modifies the magnetic free energy of an adjacent magnetic layer via magnetoelastic coupling.[9] Generally, ME composites are used for magnetic sensing,[10–12] memory and tunable radiofrequency components,[13] and antennas.[9,14] Here, a voltage sensor is presented based on a magnetization readout modulated by electric polarization.

A promising candidate for the magnetic layer are magneto-optical indicator films (MOIF) based on bismuth-substituted yttrium iron garnet (Bi:YIG). They have been implemented in many applications due to their sensitive magneto-optical magnetization readout.[15,16] MOIFs are an established platform for magnetic field sensing,[17] have been used for analysis of superconductors,[18,19] electrical steels,[20] thin magnetic films,[21] and shape-memory alloys.[22] Quantitative reconstruction of in-plane (IP)[23] and simultaneously out-of-plane (OOP) field components,[24,25] as well as current distributions,[26] has been demonstrated. Multiparametric extensions include temperature sensing via changes of saturation magnetization,[27] susceptibility[28] and combined parallel magnetic field and temperature measurement[29] and imaging.[30,31] These capabilities motivate exploring MOIFs for additional physical quantities such as voltage. Here, a perpendicular magnetic anisotropy (PMA) MOIF is employed, as they exhibit a high magneto-optical susceptibility.

In Bi:YIG systems, the magnetoelastic effect is well investigated.[32–35] Voltage-driven modulation of anisotropy and magnetization in terms of magnetic hysteresis and domain structure has been demonstrated in PMA Bi:YIG/ferroelectric heterostructures,[36–38] and also for ferrimagnetic resonance.[39] Additionally, small intrinsic magnetoelectric effects have been reported.[40] Combined with its sensitive magneto-optical magnetization readout, Bi:YIG is therefore a suitable material for investigating magneto-optical voltage sensing.

Such a magneto-optical voltage sensor is realized by mechanically coupling the Bi:YIG MOIF to a lead-zirconate-titanate (PZT) bimorph disc. An applied voltage induces bending of the disc, generating in-plane tensile or compressive stress in the magnetic layer and modulating its effective anisotropy. To achieve a high sensitivity to small perturbations, a magnetic bias field is applied to bring the sample magnetization to a critical point between magnetic in-plane saturation and an out-of-plane domain pattern.[41,42] The resulting voltage-dependent out-of-plane magnetization component under constant magnetic bias field is detected via polarimetric readout of magneto-optical Faraday rotation.[30,43] Two readout modes are demonstrated. First, a detection of AC voltages directly using the voltage induced magnetization reversal. Second, a detection of DC voltages through voltage-induced modifications of the harmonics of the magnetization response to an external applied AC field. Similar utilization of harmonics from optically read out magnetization loops has been established previously, albeit for measurement of temperature and magnetic field.[29,44,45]

## 2. Results and Discussion

### 2.1. Sensor architecture and magnetic anisotropy

The structure and operating principle of the sensor are shown in **Figure 1** (a). The sample is based on a commercially available MOIF consisting of Bi-substituted yttrium iron garnet (Bi:YIG) epitaxially grown on a single-crystalline (111) gallium gadolinium garnet (GGG) substrate with perpendicular magnetic anisotropy.[46] The Bismuth substitution enhances the magneto-optical activity, yielding a large ratio between Faraday rotation $\beta_{MO}$ to light absorption.[47] A back mirror enables operation in reflection. For perpendicular incidence of light, this results in a Faraday rotation proportional to the out-of-plane (OOP) component of magnetization $\beta_{MO} \sim M_z$. Combined with polarization optics, this enables sensitive polarimetric readout of $M_z$, especially in combination with a lock-in amplifier.[15,43] The MOIF is bonded on a commercial PZT-brass bimorph disc. The disc is fixed at two opposite points along $y$, such that an applied voltage $U_\sigma$ produces an out-of-plane bending deformation, resulting in tensile or compressive in-plane stress $\sigma_{x,U}$ in the magnetic layer. This modifies the magnetic free energy via magnetoelastic coupling and thus leads to detectable magnetization changes.

The quantitative magneto-optical hysteretic out-of-plane magnetization loop and corresponding demagnetized magneto-optical microscopy image of the sample are illustrated in Figure 1 (b) and Figure 1 (c). A maze-like domain pattern typical for thin films with perpendicular anisotropy[Hub1998] with a domain period of around $P \approx 15.6$ µm is observed. [48]

The MOIF exhibits a growth-induced magnetoelastic perpendicular anisotropy and a cubic magnetocrystalline anisotropy with easy axes along the $\langle 111 \rangle$ directions. The approximate orientation of these axes within the coordinate system is visualized in Figure 1 (d). The sample is aligned such that the direction of the voltage-induced stress, as well as any applied in-plane magnetic field, is directed along the $[\bar{1}\bar{1}2]$ direction, which is defined as the $x$-axis. As the in-plane projection of $[11\bar{1}]$ lies along $x$, the mirror symmetry of anisotropy between positive and negative $M_z$ is broken. Crucially, this results in an anisotropy dependent conversion of the average $M_z$ to an applied in-plane bias field along $x$. As components along $y$ are not considered further, the crystallographic easy axes $[1\bar{1}1]$ and $[\bar{1}11]$ are neglected. The [111] easy axis lies parallel to the uniaxial anisotropy, increasing the effective perpendicular anisotropy for small perturbations of the magnetization direction.

Now introducing the voltage-induced stress along $x$, the magnetoelastic anisotropy is modified. Within the $xz$-plane, this translates to a modulation of the effective uniaxial perpendicular anisotropy, whose energy contribution will be maximal for magnetization directions along $x$. In the absence of an external in-plane magnetic bias field, the influence of the varying uniaxial anisotropy is limited to variations in the domain wall energy, expected to alter the domain period $P$ and the corresponding magnetization loop.[49,50] Here, no voltage dependence of the domain structure or the out-of-plane magnetization loop is detected experimentally within the measurement accuracy of the experimental setup, as the relative change of uniaxial anisotropy is low.

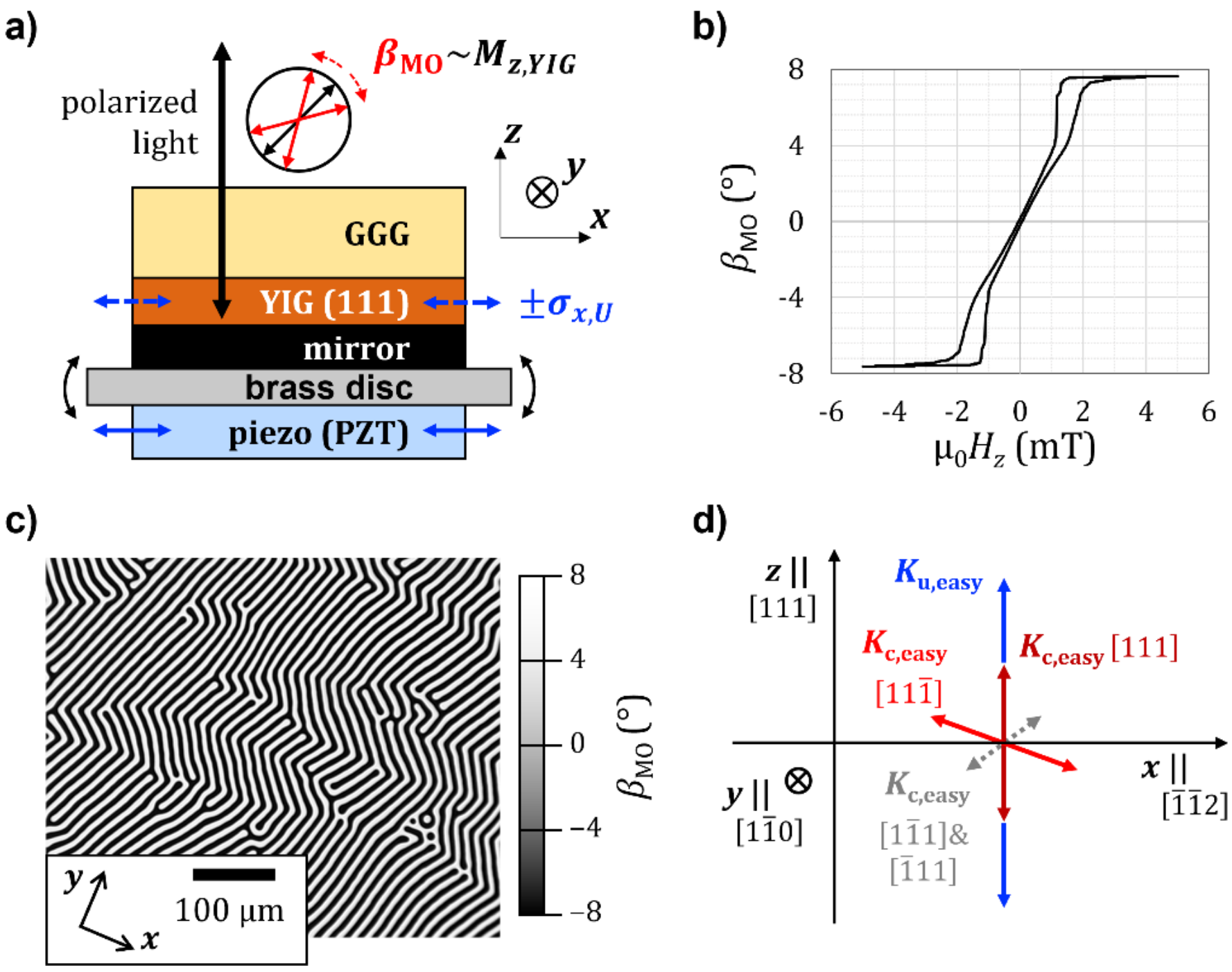

**Figure 1.** (a) Schematic sensor operating principle. Linearly polarized light is transmitted perpendicularly through the Bi:YIG magneto-optical layer, reflected back at a mirror and undergoes a Faraday rotation of the polarization axis $\beta_{MO}$ proportional to the out-of-plane magnetization component $M_z$. A PZT-brass bimorph disc is mechanically coupled to the magneto-optical stack and deforms under a voltage $U_\sigma$, generating a tensile or compressive stress $\sigma_{x,U}$ in the magnetic layer. (b) Quantitative magneto-optical out-of-plane magnetization loop of the sensor. (c) Corresponding quantitative magneto-optical microscopy image of the demagnetized magnetic domain structure. (d) Schematic of the easy axes of the crystallographic anisotropy $K_c$ and the uniaxial stress induced anisotropy $K_u$ within the $xz$-plane.

### 2.2. Voltage sensitivity via bias field activated critical domain state

To achieve a pronounced voltage response, a magnetic bias field along a hard axis within the plane in the order of the anisotropy field is applied to bring the Bi:YIG to a critical state.[41] This critical state marks the transition point between homogenous magnetization along the in-plane bias field and a stripe or bubble domain structure of perpendicularly magnetized, albeit tilted, magnetization directions.[42] According to phase theory, the energy minima corresponding to the three magnetic domain magnetization directions (“up”, “down”, “in-plane”) are nearly degenerate in this critical state.[48]The Zeeman energy compensates the perpendicular anisotropy. Together with the crystallographic easy axis along $[11\bar{1}]$ as an intermediate, this leads to a flattening of the energy landscape. An additional canted state can potentially exist as well.[51] The resulting magnetic state is susceptible to small magnetic perturbations.[41] The necessary magnetic field to reach this transition point has been shown to be dependent on the voltage of a piezoelectric plate mechanically coupled to the magnetic film, and correspondingly changes the corresponding magnetic domain pattern.[38,51]

Due to the aforementioned asymmetry introduced by the magnetocrystalline cubic anisotropy, the magnetic bias along $H_x$ necessitates an additional stabilizing out-of-plane bias component $H_z$ to reach an average $M_z$ close to zero with three domains present. For the given sample and orientation, this point is reached at $\mu_0 H_x = -30$ mT and $\mu_0 H_z = -4$ mT, which will be used in the experiments. The corresponding quantitative magneto-optical microscopy images of this state are shown in **Figure 2** (a). The applied voltage in the piezoelectric layer induces magnetic domain nucleation, annihilation and domain wall movement, as seen in the comparison to Figure 2 (b).

Both images are acquired from the same field of view as Figure 1 (c). In contrast to the unbiased case, local variations in the domain pattern become visible in this critical state. Defects with their corresponding stress fields and gradients of inhomogeneity across the sample can be clearly identified. The increased visibility of spatial inhomogeneity demonstrates the strongly enhanced susceptibility of the magnetization to the local distribution of anisotropy and thus stress in this bias-field activated high sensitivity state.

Three distinct domains are identified. Normalizing by the Faraday rotation in saturation $\beta_{\mathrm{MO,sat}}$, the magnetization cosines $m_z = \beta_{\mathrm{MO}}/\beta_{\mathrm{MO,sat}} = M_z/M_{\mathrm{sat}}$ of the stripes and bubbles of the “up” and “down” domains are measured to be significantly tilted at $m_{z,\mathrm{up}} \approx +11 \pm 3$ % and $m_{z,\mathrm{down}} \approx -38 \pm 3$ %. The third “in-plane” domain, incorporating the largest portion of the field of view, is roughly $m_{z,\mathrm{IP}} \approx -3 \pm 1$ %.

Since the voltage-induced stress alters the effective perpendicular anisotropy, the critical field to achieve the aforementioned critical state changes, as this field, in first approximation according to phase theory, corresponds directly to the effective anisotropy field.[42] Consequently, an increasing voltage leads to growth of the “down” domain area in favor of the “in-plane” domains. It stabilizes the “out-of-plane” domains and correspondingly increases the effective perpendicular anisotropy. That means rising voltage leads to increased tensile stress, as the magnetostriction constant is negative at roughly $\lambda_{111} = -1.1\times10^{-6}$ and $\lambda_{100} = -1.2\times10^{-6}$.[32,52] Additionally, the asymmetry of anisotropy from aligning the in-plane bias along the $[\bar{1}\bar{1}2]$ direction introduces an imbalance between the "up" and "down" domains. This is apparent in the different absolute magnitudes of $m_z$ of the "up" and "down" domains, and leads to the change of the average OOP magnetization with voltage. Changes in magnetization within the individual domains are also detected, but their effect on the average magnetization is relatively small compared to the observed magnetization reversal processes.

The resulting magneto-optical voltage loops from this effect are measured via a balanced photodiode setup at an excitation frequency $f = 1$ Hz. The differential change relative to the saturation signal is plotted in Figure 2 (c). The response includes both ferroelectric and ferrimagnetic hysteresis. The combined magnetoelectric coercivity voltage for a 10 V loop is measured as $U_{\mathrm{ME,c}} \approx 0.37$ V.

The influence of the applied voltage is further investigated by measuring out-of-plane magnetization loops, as illustrated in Figure 2 (d). There, an applied voltage induces a shift in the intersection points of the magnetization loop with the abscissa along the field axis, acting as an effective bias field $\mu_0 H_U$. This shift is the dominant contributor to the voltage-induced magneto-optical signal in the multidomain regime, recognizable by the hysteresis. The observed average magnitude of this effective bias per applied voltage is $\mu_0 H_U \cdot U_\sigma^{-1} \approx 13\ \mu\mathrm{T\ V}^{-1}$. In the regime of quasi homogenous magnetization, recognizable by the absence of hysteresis, the slope of the magnetization curve varies with voltage due to a changing relation between in-plane bias and the effective perpendicular anisotropy. Therefore, the sign of the measured voltage signal switches for positive and negative fields.

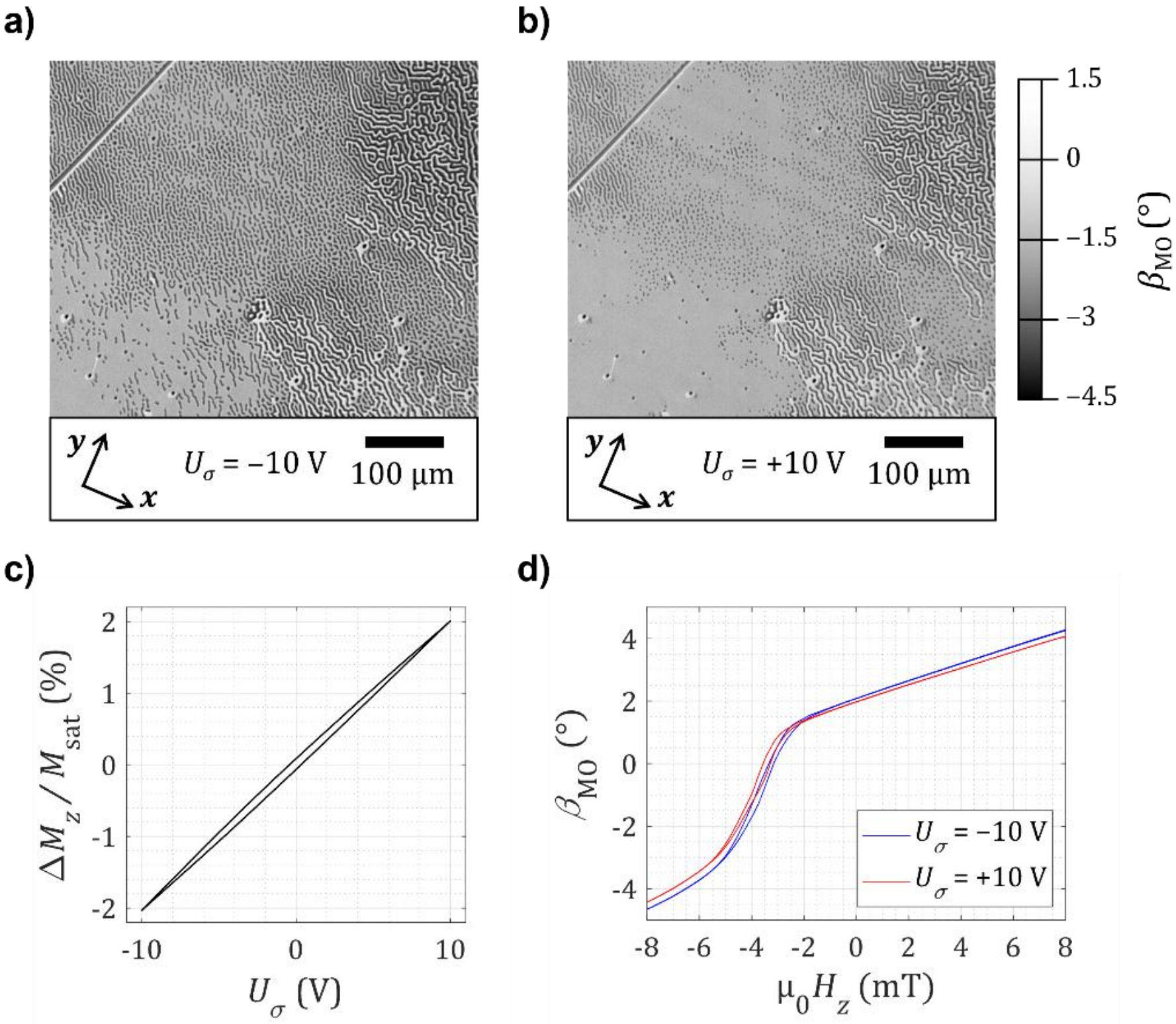

**Figure 2.** (a) & (b) Quantitative magneto-optical microscopy images of $\beta_{\mathrm{MO}}$ during an applied bias field of $\mu_0 H_x = -30$ mT and $\mu_0 H_z = -4$ mT for applied voltages of (a) $U_\sigma = -10$ V and (b) $U_\sigma = +10$ V. The applied bias field induces a critical point with three distinct magnetization directions of the magnetic domains. Defects and inhomogeneities in the sample are revealed, indicating high sensitivity to small perturbations in anisotropy, (c) Corresponding magneto-optical hysteresis loop of the perpendicular magnetization $M_z$ normalized to the saturation magnetization $M_{\mathrm{sat}}$ during an applied bias field excitation of 1 Hz measured via a balanced photodiode setup. (d) Magnetization loops to the applied out-of-plane field $H_z$ for different applied voltages $U_\sigma$ and constant applied in-plane field bias $\mu_0 H_x = -30$ mT.

### 2.3. AC sensing – voltage induced magnetization reversal

AC voltage readout is realized by exploiting the direct strain-mediated magnetization response under constant magnetic bias. To bring the Bi:YIG layer in the high-sensitivity operating state described in Section 2.2, permanent magnet ring pair is used to apply the static magnetic bias field along $x$ of 30 mT, while the bias along z is applied electromagnetically via a solenoid. An alternating voltage $U_{\sigma,\mathrm{AC}}$ at a frequency $f$ is applied to the piezoelectric layer, generating a voltage-dependent magneto-optical signal that is detected using a balanced photodiode configuration and lock-in amplification.

The amplitudes of the fundamental ($A_1$) and third harmonic ($A_3$) components of the magneto-optical signal are shown in **Figure 3** (a) and Figure 3 (b), respectively. The corresponding nested normalized minor magnetization loops from the voltage response are depicted in Figure 3 (e). The response of the fundamental harmonic is predominantly linear to the applied voltage. The slope changes at $U_{\sigma,\mathrm{AC}} \approx 12$ V. Equivalent behavior of increasing magnetic susceptibility for an applied magnetic excitation beyond a certain threshold is known and originates from the domain wall coercivity. [53[Ver2023]] This coincides with the development of the observed coercivity in the magnetization loops in Figure 3 (e). Additionally, the ferroelectric hysteresis of the PZT is included in the measurement as well. The same general principle applies to the permittivity of ferroelectrics and electric coercivity as well.[54]

The third harmonic amplitude $A_3$ is shown in Figure 3 (b) and is an indicator for non-linearity of the detected voltage response. It is at least two orders of magnitude smaller than the fundamental amplitude $A_1$. The higher harmonics quickly fall of as shown in the corresponding amplitude spectral density in Figure 3 (d). The respective total harmonic distortion is $\mathrm{THD}_{\mathrm{dB}} = -48$ dB for small voltages, which increases with voltage.

The frequency dependence of the fundamental signal amplitude is shown in Figure 3 (c). The response is highest at low frequencies and decreases gradually with increasing frequency. This is consistent with an increasing effect of domain wall coercivity with frequency of both ferrimagnetic and ferroelectrics, which then leads to the lowered magneto-optical susceptibility to voltage.[53,55] Above $f \approx 600$ Hz, the signal is dominated by a resonance of the sensor which peaks at $f \approx 800$ Hz.

The overall noise floor with peaks from the harmonics of the grid frequency at 50 Hz is visible in Figure 3 (d), as well as the peak of the resonance at $f \approx 800$ Hz. At $f = 53$ Hz the limit of detection of this setup, expressed by the equivalent noise density, is $\mathrm{END}_{\mathrm{AC}} = 0.85$ mV $\mathrm{Hz}^{-1/2}$.

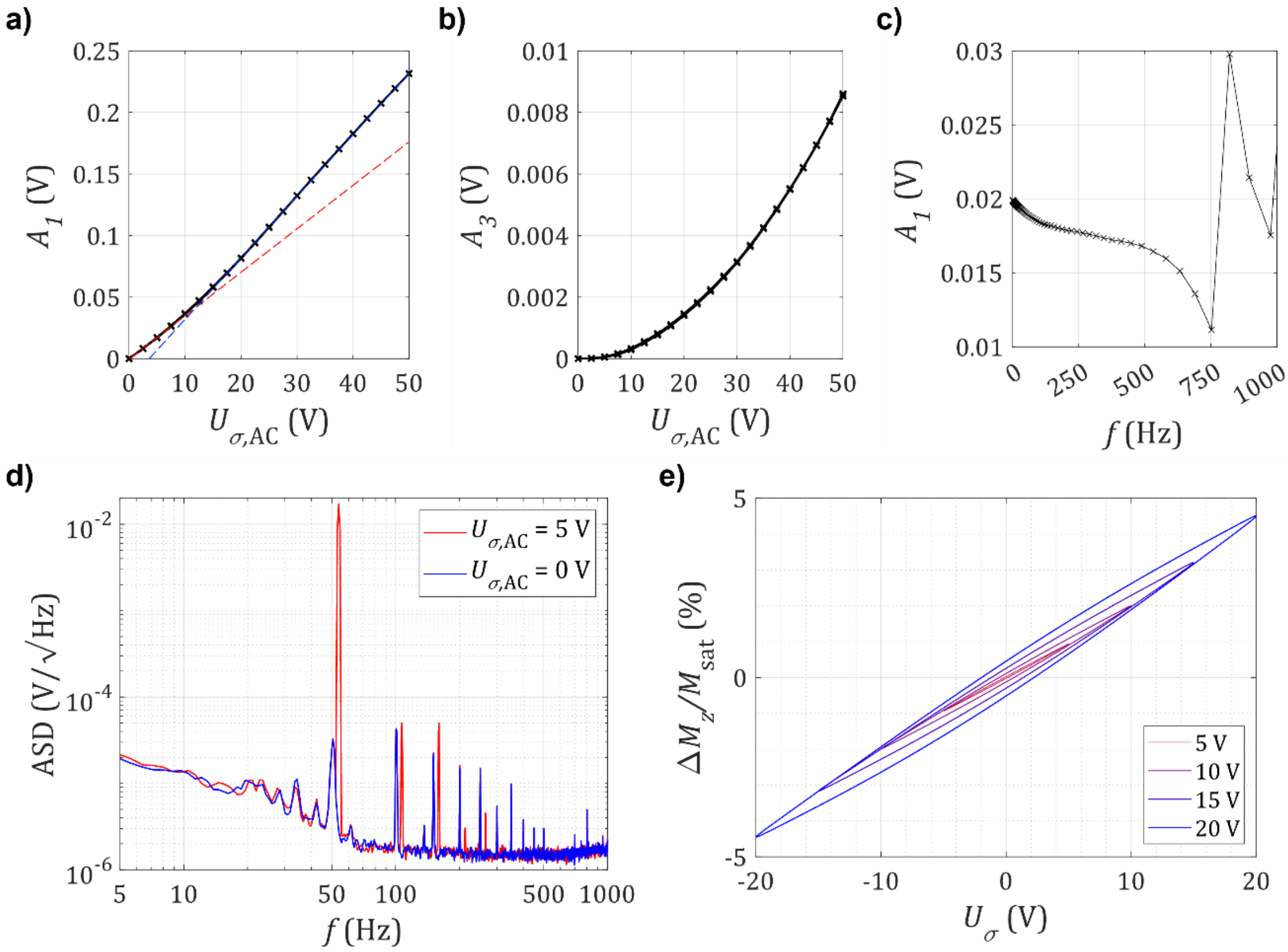

**Figure 3.** (a) Fundamental harmonic amplitude of the magneto-optical signal to the applied AC voltage in the piezoelectric layer $U_{\sigma,\mathrm{AC}}$ during an applied bias field of $\mu_0 H_x = -30$ mT and $\mu_0 H_z = -4$ mT. The red and blue dotted lines act as a visual guide for the two regions of different slope. (b) Third harmonic amplitude as a function of the voltage amplitude. (c) Fundamental signal amplitude to excitation frequency $f$ for $U_{\sigma,\mathrm{AC}} = 5$ V. (d) Amplitude spectral density ASD of the magneto-optical signal with and without an applied AC voltage in the piezoelectric layer at a frequency of $f = 53$ Hz. e) Nested minor hysteresis loops of magnetization to voltage at 53 Hz for increasing voltage amplitudes reconstructed from the 1., 2., 3. and 5. harmonic components.

### 2.4. DC sensing – voltage modulation of magnetization loop shape

Indirect voltage readout is implemented by continuous measurement of the voltage-dependent modification of the magnetization response during an OOP magnetic excitation. The corresponding magnetization loop is depicted in Figure 2 (d). A constant DC voltage $U_{\sigma,\mathrm{DC}}$ is applied to the piezoelectric layer, while the magnetization of the Bi:YIG film is periodically excited by an out-of-plane alternating magnetic field at a frequency of $f = 123$ Hz, with an amplitude of $\mu_0 H_{\mathrm{AC}} = 6$ mT and a constant bias of $\mu_0 H_{\mathrm{DC}} = -4$ mT. The resulting magneto-optical signal containing information about the hysteresis shape is detected using the same polarimetric setup with lock-in amplification as for the AC readout.[44,56] This enables direct harmonic analysis of the magnetization response for DC voltage redout.

The amplitudes of the fundamental ($A_1$), second ($A_2$), and fifth ($A_5$) harmonic components of the magneto-optical signal are shown in **Figure 4** (a)-(c) as a function of the applied DC voltage $U_{\sigma,\mathrm{DC}}$. The data is corrected for a linear temperature drift, as saturation magnetization and anisotropy, and thus the detected signal, are dependent on temperature.[44] The measurement is carried out as a voltage loop. It contains no magnetic hysteresis as the magnetization is continuously excited and a single domain state is reached in every cycle. The ferroelectric coercivity for a 10 V loop is determined as $U_{E,c} \approx 0.30$ V. Comparing Figure 4 (a) with Figure 2 (a), the observed coercivity is therefore mostly associated with the ferroelectric behavior of the piezoelectric actuator.

The fundamental harmonic represents the average magnetic susceptibility and corresponds the changing slope in Figure 2 (d). The second harmonic contains information about the asymmetry of the loop and thus contains information about the shift by the voltage induced effective bias field of the loop. It also changes significantly with the applied $H_z$.

The higher harmonics describe further nonlinearities of the loop. Their behavior can be extracted from the amplitude spectral density depicted in Figure 4 (d). It additionally contains the noise floor. A peak at 1517 Hz together with side peaks in 246 Hz intervals (double the excitation frequency) is observed. These originate from magnetoelectric coupling of the magnetic excitation to resonating mechanical vibrations.

The evolution of the harmonics to changes in voltage can be extracted from the absolute difference in the amplitude spectral densities shown in Figure 4 (e). As the noise floor is approximately constant, the height of these differential peaks indicates the average sensitivity within this voltage regime. The corresponding average equivalent voltage noise densities are for the fundamental signal $\mathrm{END}_{\mathrm{DC,A1}} = 6.7\ \mathrm{mV\ Hz^{-1/2}}$, for the second harmonic $\mathrm{END}_{\mathrm{DC,A2}} = 13\ \mathrm{mV\ Hz^{-1/2}}$, and for the fifth harmonic $\mathrm{END}_{\mathrm{DC,A5}} = 18\ \mathrm{mV\ Hz^{-1/2}}$. Adapting the methodology of Klingbeil et al.[44] using ratios of different harmonic amplitudes, the readout can be made insensitive to fluctuations of e.g. the absolute optical intensity.

For DC measurements, this indirect approach is significantly more accurate than the direct readout described in **2.3** due to 1/*f* noise. The limiting factor is the hysteresis, which introduces an ambiguity up to the electric coercivity, which corresponds to with a range of 10 V to 3 % of the full scale.

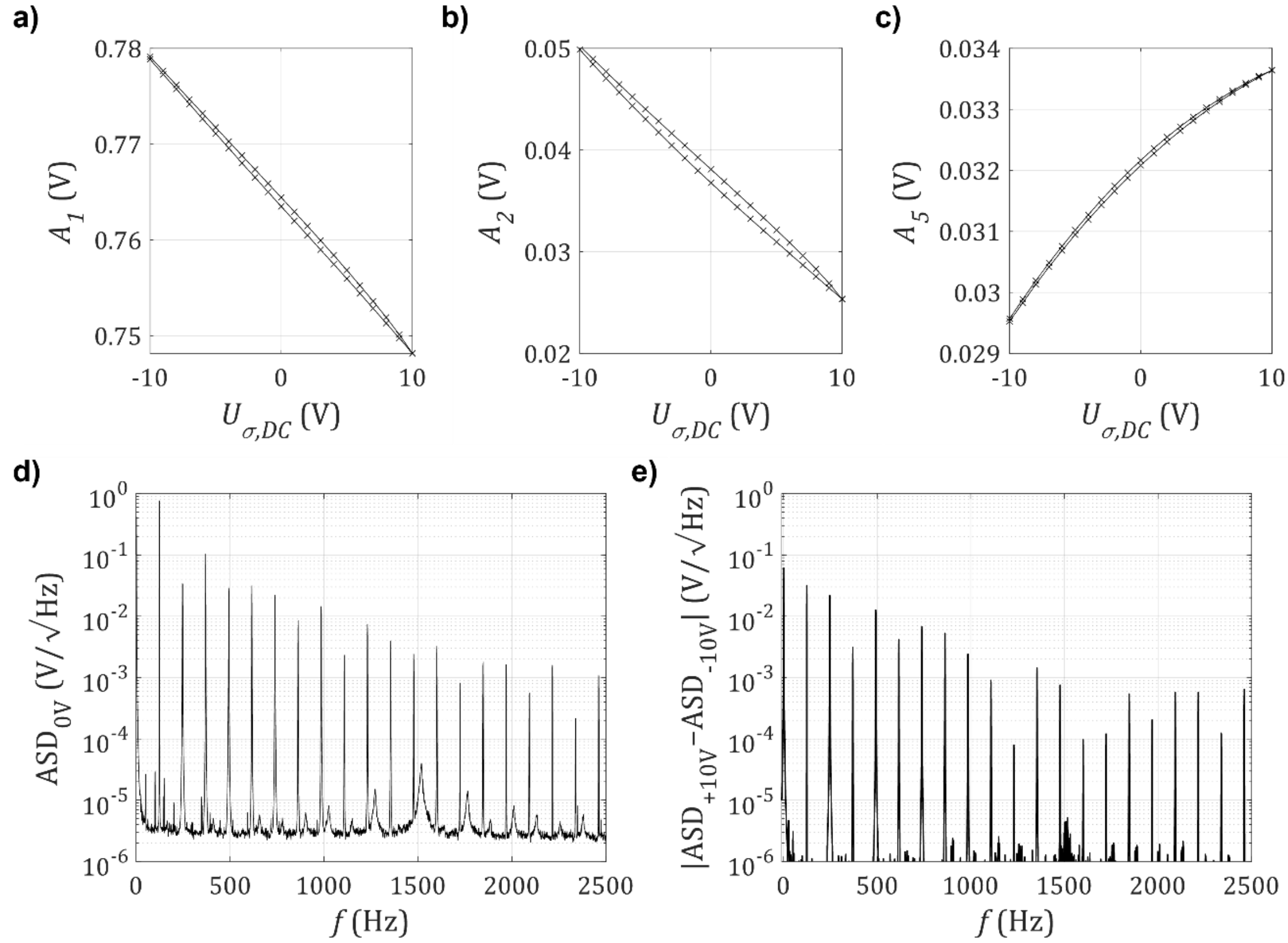

**Figure 4.** (a) Fundamental, (b) second and (c) third harmonic amplitude of the magneto-optical signal during an applied magnetic excitation $\mu_0 H_{AC}$ = 6 mT at $f$ = 123 Hz at a constant magnetic bias of $\mu_0 H_{DC}$ = −4 mT as a function of a constant applied piezoelectric voltage $U_{\sigma,DC}$ measured as a voltage loop. The data is corrected for a linear temperature drift. (d) Corresponding amplitude spectral density at $U_{\sigma,DC}$ = 0 V. (e) absolute difference of the amplitude spectral densities between $U_{\sigma,DC}$ = 10 V and $U_{\sigma,DC}$ = −10 V.

## 2.5. Discussion

For high-voltage applications, the presented voltage sensor concept can be combined with passive voltage scaling elements. As discussed for piezoelectric optical voltage sensors by Edmunds et al.,[7] capacitive dividers based on low-capacitance components enable scaling of high voltages into a range suitable for piezoelectric actuation while preserving galvanic isolation.

The presented sensing scheme is a proof of concept, and as such it could be further refined and optimized in terms of minimization, sensitivity and robustness in future research. The mechanical coupling could be enhanced by omitting the intermediate bimorph disc, such that Bi:YIG and piezoelectric film layers are adjacent. [37] This is expected to enhance strain transfer and increase the achievable sensor bandwidth. In addition, the magneto-optical readout can be integrated using fiber-optic approaches, as established for garnet-based magnetic sensing.[57] Robustness of the sensor could be achieved using multiparametric readout schemes, a topic that has gained increasing attention in research with optical sensors.[58] Multiparametric sensing can be realized via already established methods for MOIF-based detection.[29,31]

In this example, a parallel acquisition of temperature can be achieved by using multiple harmonic components as described in **2.5** and a multidimensional optimization solved for example via machine learning. Such a simultaneous measurement could directly compensate for the temperature drift. An analogous approach could be applied to compensate magnetic field variations as well.

## 3. Conclusion

This work introduces a novel magneto-optical voltage sensor concept that couples a Bi:YIG magneto-optical indicator film with a piezoelectric bending actuator. An applied voltage generates stress in the garnet layer and modifies the magnetic free energy via magnetoelastic coupling. A pronounced response occurs when the (111)-oriented magnetic film is brought into a critical high sensitivity state. Using a magnetic IP bias along the primary stress direction parallel to the $[\bar{1}\bar{1}2]$ direction, the asymmetry introduced by the crystallographic anisotropy introduces a dependence of the average magnetization to the perpendicular anisotropy. An OOP bias then stabilizes the critical state between homogenous in-plane magnetization and a domain structure of perpendicular domains, achieving high sensitivity to small perturbations of anisotropy and thus voltage.

Two readout modes are demonstrated. In AC sensing, the direct effect of voltage on the average magnetization is exploited. The optical signal follows the applied voltage in a predominantly linearly fundamental response with $\mathrm{THD_{dB}} = -48$ dB. The equivalent voltage noise density reaches $\mathrm{END_{AC}} = 0.85$ mV $\mathrm{Hz^{-1/2}}$ at 53 Hz. In indirect DC sensing, the shape of the magnetization loop is read out as the harmonics of the signal. There, the modifications of the loop by the voltage are detected. Using the change of the first harmonic or average magneto-optical susceptibility, an equivalent voltage noise density of down to $\mathrm{END_{DC}} = 6.7$ mV $\mathrm{Hz^{-1/2}}$ is measured. It is limited by the electric coercivity of the ferroelectric hysteresis $U_C \approx 0.3$ V.

Overall, the presented results lay a foundation for strain-mediated magneto-optical voltage sensing. At present, the voltage readout is intrinsically linked to magnetic bias conditions and temperature-dependent magnetic properties, though it exhibits sufficiently low noise for practical high voltage applications. As a new measurement approach, multiple routes for optimization are available. In particular, multiparametric sensing strategies provide a pathway to utilize dependencies on magnetic field and temperature, allowing the concept to evolve toward a competitive voltage measurement scheme.

## 4. Experimental Section/Methods

*Materials*:

The sample is based on a commercially available MOIF consisting of Bi- and Al- substituted yttrium iron garnet (Bi:YIG) epitaxially grown on a single-crystalline (111) gallium gadolinium (GGG) substrate.[46] The Bismuth substitution enhances the magneto-optical activity, yielding a large Faraday rotation $\beta_{MO}$ at low absorption.[47] The MOIF exhibits a Néel temperature of $T_c = 140$ °C and a maximal Faraday rotation in saturation of $\beta_{MO,sat} = 7.8°$ at $\lambda = 554$ nm at room temperature. The MOIF is bonded on the commercial PZT-brass bimorph disc using mounting wax. The capacitance of the piezoelectric component is $C$ = 20 nF.

*Polarimetric Balanced Photodiode Setup*:

Both AC and DC demonstrated voltage measurement schemes are measured using a dual quadrature polarimetric setup.[28,43] Light from an LED source with $\lambda$ = 554 nm is coupled into a multimode optical fiber, collimated, linearly polarized and directed onto the sample surface through a polarization preserving beamsplitter. The illuminated area has a diameter of approximately 3 mm. The reflected Faraday-rotated light is redirected by the beamsplitter into the detection arm and separated into two orthogonal polarization channels by a Wollaston prism. The beams are detected by a balanced photodiode. The resulting signal is in approximation proportional to $\beta_{\mathrm{MO}}$. The signal is processed by a lock-in amplifier referenced either to the applied voltage (AC) or the applied magnetic excitation (DC).

To ensure the signal is purely magnetic and does not originate from alterations of the reflection angle due to physical movement of the sensor, the optical signal amplitude during voltage excitation is minimized during OOP saturation. The remaining mechanical signal is at least two orders of magnitude smaller than all other reported fundamental signal amplitudes in this work, and are thus considered negligible.

The $x$-component of applied magnetic field is applied with a pair of ring-shaped permanent magnets. For any applied fields along $z$ a coil in Helmholtz configuration is used.

*Polarimetric Microscopy*:

Magneto-optical images are acquired using wide-field magneto-optical microscopy with Köhler type illumination using a white light source, a 5x Objective and aperture in the back focal plane in polar configuration.[59,60] Quantified detection is realized via a Stokes polarization camera.[61] Any magnetic field is applied with a three-dimensional vector electromagnet. For setup details and quantitative calibration see Path et al.[30]

## CRediT authorship contribution statement

**M. Path**: Writing – review & editing, Writing – original draft, Visualization, Validation, Investigation, Conceptualization, Formal analysis, Software, Methodology. **J. McCord**: Writing – review & editing, Validation, Supervision, Project administration, Funding acquisition, Conceptualization, Resources.

## Conflict of Interest

The authors have no conflicts to disclose.

### Acknowledgements

We acknowledge funding by the German Research Foundation (DFG) - MC 9/20-2.

**Data Availability Statement**

The datasets generated and analyzed during the current study are available from the corresponding author on reasonable request.

**References**

1. Wetula, A., Bień, A. & Parekh, M. New Sensor for Medium- and High-Voltage Measurement. *Energies* **14,** 4654; 10.3390/en14154654 (2021).
2. Bi, L. & Li, H. An Overview of Optical Voltage Sensor. In *2012 International Conference on Computer Science and Electronics Engineering* (IEEE2012), pp. 197–201.
3. Yoshino, T., Kurosawa, K., Itoh, K. & Ose, T. Fiber-Optic Fabry-Perot Interferometer and its Sensor Applications. *IEEE Trans. Microwave Theory Techn.* **30,** 1612–1621; 10.1109/TMTT.1982.1131298 (1982).
4. Zeng, R., Wang, B., Yu, Z. & Chen, W. Design and application of an integrated electro-optic sensor for intensive electric field measurement. *IEEE Trans. Dielect. Electr. Insul.* **18,** 312–319; 10.1109/TDEI.2011.5704523 (2011).
5. Mitsui, T., Hosoe, K., Usami, H. & Miyamoto, S. Development of Fiber-Optic Voltage Sensors and Magnetic-Field Sensors. *IEEE Trans. Power Delivery* **2,** 87–93; 10.1109/TPWRD.1987.4308077 (1987).
6. Da Allil, R. C. S. B. & Werneck, M. M. Optical High-Voltage Sensor Based on Fiber Bragg Grating and PZT Piezoelectric Ceramics. *IEEE Trans. Instrum. Meas.* **60,** 2118–2125; 10.1109/TIM.2011.2115470 (2011).
7. Edmunds, J. L., Sonmezoglu, S., Martens, J., Meier, A. von & Maharbiz, M. M. Optical voltage sensor based on a piezoelectric thin film for grid applications. *Opt. Express, OE* **29,** 33716–33727; 10.1364/OE.437915 (2021).
8. Dante, A., Bacurau, R. M., Spengler, A. W., Ferreira, E. C. & Dias, J. A. S. A Temperature-Independent Interrogation Technique for FBG Sensors Using Monolithic Multilayer Piezoelectric Actuators. *IEEE Trans. Instrum. Meas.* **65,** 2476–2484; 10.1109/TIM.2016.2594021 (2016).
9. Cheng, Y., Peng, B., Hu, Z., Zhou, Z. & Liu, M. Recent development and status of magnetoelectric materials and devices. *Phys. Lett. A* **382,** 3018–3025; 10.1016/j.physleta.2018.07.014 (2018).
10. Bichurin, M. *et al.* Magnetoelectric Magnetic Field Sensors: A Review. *Sensors* **21,** 6232; 10.3390/s21186232 (2021).
11. Viehland, D., Wuttig, M., McCord, J. & Quandt, E. Magnetoelectric magnetic field sensors. *MRS Bull.* **43,** 834–840; 10.1557/mrs.2018.261 (2018).
12. Elzenheimer, E. *et al.* Quantitative Evaluation for Magnetoelectric Sensor Systems in Biomagnetic Diagnostics. *Sensors* **22**; 10.3390/s22031018 (2022).
13. Hu, J.-M., Nan, T., Sun, N. X. & Chen, L.-Q. Multiferroic magnetoelectric nanostructures for novel device applications. *MRS Bull.* **40,** 728–735; 10.1557/mrs.2015.195 (2015).
14. Ge, L. & Luk, K. M. A Low-Profile Magneto-Electric Dipole Antenna. *IEEE Trans. Antennas Propagat.* **60,** 1684–1689; 10.1109/TAP.2012.2186260 (2012).
15. Dorosinskiy, L. & Sievers, S. Magneto-Optical Indicator Films: Fabrication, Principles of Operation, Calibration, and Applications. *Sensors* **23,** 4048; 10.3390/s23084048 (2023).
16. Arsad, A. Z., Zuhdi, A. W. M., Ibrahim, N. B. & Hannan, M. A. Recent Advances in Yttrium Iron Garnet Films: Methodologies, Characterization, Properties, Applications, and Bibliometric Analysis for Future Research Directions. *Appl. Sci.* **13,** 1218; 10.3390/app13021218 (2023).

17. Rochford, K. B., Rose, A. H. & Day, G. W. Magneto-optic sensors based on iron garnets. *IEEE Trans. Magn.* **32,** 4113–4117; 10.1109/LEOS.1996.565218 (1996).
18. Shaw, G. *et al.* Quantitative magneto-optical investigation of superconductor/ferromagnet hybrid structures. *Rev. Sci. Instrum.* **89,** 23705; 10.1063/1.5016293 (2018).
19. Koblischka, M. R. & Wijngaarden, R. J. Magneto-optical investigations of superconductors. *Supercond. Sci. Technol.* **8,** 199–213; 10.1088/0953-2048/8/4/002 (1995).
20. Schäfer, R., Soldatov, I. & Arai, S. Power frequency domain imaging on Goss-textured electrical steel. *J. Magn. Magn. Mater.* **474,** 221–235; 10.1016/j.jmmm.2018.10.100 (2019).
21. Bennett, L. H. *et al.* Magneto-optical indicator film observation of domain structure in magnetic multilayers. *Appl. Phys. Lett.* **66,** 888–890; 10.1063/1.113421 (1995).
22. Neudert, A. *et al.* Magnetic Domains and Twin Boundary Movement of NiMnGa Magnetic Shape Memory Crystals. *Adv. Eng. Mater.* **14,** 601–613; 10.1002/adem.201200074 (2012).
23. Rogachev, A. E. *et al.* Vector magneto-optical sensor based on transparent magnetic films with cubic crystallographic symmetry. *Appl. Phys. Lett.* **109,** 162403; 10.1063/1.4964887 (2016).
24. Sakaguchi, H. *et al.* 3D Magnetic Field Vector Measurement by Magneto-Optical Imaging. *J. Magn. Soc. Jpn.* **46,** 37–41; 10.3379/msjmag.2203R002 (2022).
25. Patterson, W. C., Garraud, N., Shorman, E. E. & Arnold, D. P. A magneto-optical microscope for quantitative measurement of magnetic microstructures. *Rev. Sci. Instrum.* **86,** 94704; 10.1063/1.4930178 (2015).
26. Arakelyan, S. *et al.* Direct current imaging using a magneto-optical sensor. *Sens. Actuators, A* **238,** 397–401; 10.1016/j.sna.2016.01.002 (2016).
27. Kustov, M., Grechishkin, R., Gusev, M., Gasanov, O. & McCord, J. A Novel Scheme of Thermographic Microimaging Using Pyro-Magneto-Optical Indicator Films. *Adv. Mater.* **27,** 5017–5022; 10.1002/adma.201501859 (2015).
28. Klingbeil, F., Stölting, S. D. & McCord, J. Sensing of temperature through magnetooptical domain wall susceptibility. *Appl. Phys. Lett.* **118,** 92403; 10.1063/5.0037128 (2021).
29. Path, M. P., Vogel, M. & McCord, J. Multiparametric robust sensing via readout of characteristic magnetization loops. *Sci. Rep.* **16,** 8148; 10.1038/s41598-026-42763-x (2026).
30. Path, M. P. & McCord, J. Quantitative 3D magnetic vector field imaging and thermal sensing with magneto-optical indicator films. *J. Appl. Phys.* **138**; 10.1063/5.0301604 (2025).
31. Lee, H., Jeon, S., Friedman, B. & Lee, K. Simultaneous imaging of magnetic field and temperature distributions by magneto optical indicator microscopy. *Sci. Rep.* **7,** 43804; 10.1038/srep43804 (2017).
32. Hansen, P., Witter, K. & Tolksdorf, W. Magnetic and magneto-optic properties of lead- and bismuth-substituted yttrium iron garnet films. *Phys. Rev. B* **27,** 6608–6625; 10.1103/PhysRevB.27.6608 (1983).
33. Nistor, I., Krafft, C., Rojas, R. & Mayergoyz, I. D. Measurement of the Magnetostriction Constant of Bi-Doped Garnets by Optical Observation of Stress-Induced Stripe Domains. *IEEE Trans. Magn.* **40,** 2832–2834; 10.1109/TMAG.2004.829015 (2004).
34. Nakamura, A. & Sugiura, Y. Magnetostriction of Yttrium Iron Garnet. *J. Phys. Soc. Jpn.* **15,** 1704; 10.1143/JPSJ.15.1704 (1960).
35. Linchevskyi, I. V. Reverse Magnetomechanical Effect Measurement in Magnetooptical Films at Nonmagnetic Substrate Bending Deformation. *IEEE Trans. Magn.* **52,** 1–4; 10.1109/TMAG.2015.2509027 (2016).
36. Gross, M. J. *et al.* Voltage modulated magnetic anisotropy of rare earth iron garnet thin films on a piezoelectric substrate. *Appl. Phys. Lett.* **121**; 10.1063/5.0128842 (2022).

37. Misba, W. A. *et al.* Strain Mediated Voltage Control of Magnetic Anisotropy and Magnetization Reversal in Bismuth-Substituted Yttrium Iron Garnet Films and Mesostructures. *ACS Appl. Mater. Interfaces* **17,** 65956-65955; 10.1021/acsami.5c14761 (2025).
38. Sohatsky, V., Kostuk, A. & Savytsky, M. Reorientation of Magnetization in Single Crystalline Yttrium Iron Garnet Film under Mechanical Strain. *Solid State Phenom.* **230,** 259–263; 10.4028/www.scientific.net/SSP.230.259 (2015).
39. Pettiford, C., Dasgupta, S., Lou, J., Yoon, S. D. & Sun, N. X. Bias Field Effects on Microwave Frequency Behavior of PZT/YIG Magnetoelectric Bilayer. *IEEE Trans. Magn.* **43,** 3343–3345; 10.1109/TMAG.2007.893790 (2007).
40. Logginov, A. S., Meshkov, G. A., Nikolaev, A. V. & Pyatakov, A. P. Magnetoelectric control of domain walls in a ferrite garnet film. *JETP Lett.* **86,** 115–118; 10.1134/S0021364007140093 (2007).
41. Fichtner, R., Hubert, A. & Grimm, H. The detection of magnetic inclusions in the surface of a non-magnetic ceramic using the critical state of a garnet film. *13. Intern. Conf. Magnetic Films and Surfaces,* 423–424 (1991).
42. Hubert, A., Malozemoff, A. P. & DeLuca, J. C. Effect of cubic, tilted uniaxial, and orthorhombic anisotropies on homogeneous nucleation in a garnet bubble film. *J. Appl. Phys.* **45,** 3562–3571; 10.1063/1.1663818 (1974).
43. Deeter, M. N., Rose, A. H. & Day, G. W. Fast, sensitive magnetic-field sensors based on the Faraday effect in YIG. *J. Lightwave Technol.* **8,** 1838–1842; 10.1109/50.62880 (1990).
44. Klingbeil, F., Path, M. P. & McCord, J. Robust magneto-optical temperature sensing using harmonics of magnetic loop characteristics. *Rev. Sci. Instrum.* **97,** 14901; 10.1063/5.0301082 (2026).
45. García, R., Blanco, E. & Domínguez, M. Development of a magneto-optical sensor prototype to measure current by means of the induced magnetic field. *Sens. Actuators, A* **249,** 231–241; 10.1016/j.sna.2016.08.010 (2016).
46. matesy GmbH. Magneto-optical sensors. Magneto-Optical Visualization of Magnetic Fields. Available at https://matesy.de/en/products/magneto-optical-sensors (2026).
47. Aichele, T., Lorenz, A., Hergt, R. & Görnert, P. Garnet layers prepared by liquid phase epitaxy for microwave and magneto‐optical applications ‐ a review. *Cryst. Res. Technol.* **38,** 575–587; 10.1002/crat.200310071 (2003).
48. Hubert, A. & Schäfer, R. *Magnetic domains. The analysis of magnetic microstructures* (Springer, Berlin, Heidelberg, 1998).
49. Kittel, C. Physical Theory of Ferromagnetic Domains. *Rev. Mod. Phys.* **21,** 541; 10.1103/RevModPhys.21.541 (1949).
50. Bogdanov, A. N. & Yablonskii, D. A. Theory of the domain strcuture in ferrimagnets. *Sov. Phys. Solid State* **22,** 399–403 (1980).
51. Zavislyak, I. V., Sohatsky, V. P., Popov, M. A. & Srinivasan, G. Electric-field-induced reorientation and flip in domain magnetization and light diffraction in an yttrium-iron-garnet/lead-zirconate-titanate bilayer. *Phys. Rev. B* **87**; 10.1103/PhysRevB.87.134417 (2013).
52. Hansen, P., Witter, K. & Tolksdorf, W. Magnetic and magneto‐optic properties of bismuth‐ and aluminum‐substituted iron garnet films. *J. Appl. Phys.* **55,** 1052–1061; 10.1063/1.333187 (1984).
53. Vértesy, G. Coercive Properties of Magnetic Garnet Films. *Crystals* **13,** 946; 10.3390/cryst13060946 (2023).

54. Damjanovic, D., Bharadwaja, S. & Setter, N. Toward a unified description of nonlinearity and frequency dispersion of piezoelectric and dielectric responses in Pb(Zr,Ti)O3. *Mater. Sci. Eng., B* **120,** 170–174; 10.1016/j.mseb.2005.02.011 (2005).
55. Lente, M. H., Picinin, A., Rino, J. P. & Eiras, J. A. 90° domain wall relaxation and frequency dependence of the coercive field in the ferroelectric switching process. *J. Appl. Phys.* **95,** 2646–2653; 10.1063/1.1645980 (2004).
56. Willcock, S. & Tanner, B. Harmonic analysis of B-H loops. *IEEE Trans. Magn.* **19,** 2265–2270; 10.1109/TMAG.1983.1062588 (1983).
57. Sohlström, H. Fibre Optic Magnetic Field Sensing Utilizing Iron Garnet Materials. Dissertation. Royal Institute of Technology, 1993.
58. Ma, G. *et al*. Optical sensors for power transformer monitoring: A review. *High Voltage* **6,** 367–386; 10.1049/hve2.12021 (2021).
59. Schäfer, R. & McCord, J. Magneto-Optical Microscopy. In *Magnetic Measurement Techniques for Materials Characterization*, edited by V. Franco & B. Dodrill. 1st ed. (Springer International Publishing; Imprint Springer, Cham, 2021), pp. 171–229.
60. McCord, J. Progress in magnetic domain observation by advanced magneto-optical microscopy. *J. Phys. D: Appl. Phys.* **48,** 333001; 10.1088/0022-3727/48/33/333001 (2015).
61. Vedel, M., Lechocinski, N. & Breugnot, S. Compact and robust linear Stokes polarization camera. *EPJ Web of Conferences* **5,** 1005; 10.1051/epjconf/20100501005 (2010).